\documentclass[numberedappendix]{emulateapj}
\usepackage{apjfonts}

\newcommand{\gtsima}{$\; \buildrel > \over \sim \;$}
\newcommand{\ltsima}{$\; \buildrel < \over \sim \;$}
\newcommand{\simgt}{\lower.5ex\hbox{\gtsima}}
\newcommand{\simlt}{\lower.5ex\hbox{\ltsima}}
\newcommand{\himpc}{{\hbox {$\,h^{-1}$}{\rm Mpc}} }

\newcommand{\bfm}[1]{{\mbox{\boldmath $#1$}}}
\newcommand{\sbfm}[1]{{\mbox{\scriptsize\boldmath $#1$}}}

\newcommand{\mOmega}{{\mit \Omega}}
\newcommand{\mLambda}{{\mit \Lambda}}
\newcommand{\mGamma}{{\mit \Gamma}}

\begin{document}

%\renewcommand{\thefootnote}{\fnsymbol{footnote}}
%\renewcommand{\theequation}{\mbox{\rm
%{\arabic{section}.\arabic{equation}}}} 
%\renewcommand{\theequation}{\mbox{\rm {\arabic{equation}}}} 

%MMMMMMMMMMMMMMMMMMMMMMMMMMMMMMMMMMMMMMMMMMMMMMMMMMMMM

\title{Correlation Function in Deep Redshift Space as a Cosmological
Probe}

\author{Takahiko Matsubara}
\affil{Department of Physics and Astrophysics, 
	Nagoya University,
	Chikusa, Nagoya 464-8602, Japan}

\begin{abstract}
Recent development of galaxy surveys enables us to investigate the
deep universe of high redshift. We quantitatively present the physical
information extractable from the observable correlation function in
deep redshift space in a framework of the linear theory. The
correlation function depends on the underlying power spectrum,
velocity distortions, and the Alcock-Paczy\'nski (AP) effect. The
underlying power spectrum is sensitive to the constituents of matters
in the universe, the velocity distortions are sensitive to the galaxy
bias as well as the amount of total matter, and the Alcock-Paczy\'nski
effect is sensitive to the dark energy components. Measuring the dark
energy by means of the baryonic feature in the correlation function is
one of the most interesting applications. We show that the ``baryon
ridge'' in the correlation function serves as a statistically circular
object in the AP effect.  In order to sufficiently constrain the dark
energy components, the redshift range of the galaxy survey should be
as broad as possible. The survey area on the sky should be smaller at
deep redshifts than at shallow redshifts to keep the number density as
dense as possible. We illustrate an optimal survey design that are
useful in cosmology. Assuming future redshift surveys of $z\simlt 3$
which are within reach of the present-day technology, achievable error
bounds on cosmological parameters are estimated by calculating the
Fisher matrix. According to an illustrated design, the equation of
state of dark energy can be constrained within $\pm 5\%$ error
assuming that the bias is unknown and marginalized over. Even when all
the other cosmological parameters should be simultaneously determined,
the error bound for the equation of state is up to $\pm 10\%$.
\end{abstract}

%MMMMMMMMMMMMMMMMMMMMMMMMMMMMMMMMMMMMMMMMMMMMMMMMMMMMMMMMMM


\setcounter{equation}{0}
\section{
Introduction
\label{sec1}
}

Observing galaxy clustering in deep redshift surveys is one of the most
direct way to probe the evolution and structure of our universe
itself. Recent advances in observational technologies of deep surveys
are spectacular, and enormous amount of information on deep universe
must be flooded in near future. To understand the cosmological meaning
of galaxy clustering in deep universe, theoretical analysis of the
clustering of observed galaxies are necessary.

The observable objects in the deep redshift surveys are fainter than
the shallow surveys, and the sampling number density per comoving
volume is smaller if we adopt selection criteria by apparent magnitude
of galaxies. When one is interested in linear clustering regime, where
complex nonlinear dynamics is not relevant, the mean separation of the
objects should be at least of order $10\himpc$. Recent advances of
redshift surveys, such as the Deep Extragalactic Evolutionary Probe
survey \citep[DEEP;][]{davis03}, and the Sloan Digital Sky Survey
\citep[SDSS;][]{york00}, etc.~can map such fainter objects with
sufficient qualities. By an efficient color-selection technique, the
sampling number density can be improved.

The deep redshift surveys, if the sampling density is large enough,
have many advantages. First, the observable volume at deep redshifts
is much larger than that at shallow redshifts, simply because the
available volume is large in distant space. Second, the nonlinear
clustering properties are less important at a fixed scale in deep
surveys, because the universe is younger in deep universe and there is
less time for the density fields to evolve nonlinearities. Third,
while the observations of the nearby universe can map only the
snapshot of the recent universe, the observations of the deep universe
reveal the evolutionary properties of the universe. This is an
indispensable information to establish a consistent picture of the
whole universe.

The measurements of the cosmic microwave background, such as the
Wilkinson Microwave Anisotropy Probe (WMAP) satellite
\citep{bennett03}, provide the information of the universe at very
high redshift, $z \simeq 1100$. On the other hand, the traditional
galaxy redshift surveys, such as the Las Campanas Redshift Survey
\citep[LCRS,][]{shectman96}, IRAS Point Source Catalog Redshift Survey
\citep[PSCz,][]{saunders00}, and the AAT two-degree field galaxy
redshift survey \citep[2dFGRS,][]{colless01} reveal the recent
universe, $z \simlt 0.2$. Systematic investigations of the universe
between these redshifts are undoubtedly promising ways to unveil the
evolutionary feature of the universe. Especially, the galaxy redshift
surveys of redshift range of $z = 0.2$--$3.0$ can be attainable by
present-day technology. The Kilo-Aperture Optical Spectrograph (KAOS)
project\footnote{http://www.noao.edu/kaos/} is proposed to survey
galaxies at those redshifts with sufficient density, using one of the
existing Gemini telescopes.

One of the most significant feature of the universe one can probe by
such deep redshift surveys is the nature of the dark energy component
of the universe. The dark energy component does not have important
contributions at very high redshifts. The effects of the dark energy
can be probed by studying how the recent universe evolves with
redshifts. One of the striking evidence of the dark energy was
provided by the luminosity distance--redshift relation of type Ia
supernovae \citep{riess98,perlmutter99} at redshifts $z \simlt 1$. The
galaxy redshift surveys had not been considered to provide constraints
on the dark energy, since the galaxy surveys had been restricted to
the low-redshift universe at $z \simlt 0.2$. The ongoing and future
galaxy surveys will break this limitation and the evolutionary
features of the universe, including dark energy properties, will be
searched by galaxy surveys. A machinery of probing the dark energy by
galaxy surveys is based on the extended Alcock-Paczy\'nski (AP)
effects \citep{alcock79,ballinger96,matsubara96} on galaxy clustering
in redshift space. Recently, several authors investigate the
feasibility of detecting the nature of the dark energy from the
extended AP effects of the correlation function \citep{matsubara03}
and the power spectrum \citep{blake03,seo03,hu03,linder03}.

Theoretical predictions of the extended AP effect are easier when the
power spectrum is used, because the dependences of the observed power
spectrum with AP effect on cosmological parameters are not complicated
once the dependence on the mass power spectrum is known. Including
simultaneously the peculiar velocity effect \citep{kaiser87} and the
AP effect on the power spectrum requires only simple procedures to
implement \citep{ballinger96}. However, observational determination of
the reliable power spectrum requires a homogeneous sample, since the
observed spectrum is a convolution of the survey geometry and the real
density spectrum. On the other hand, the correlation function is more
straightforward to observationally determine. Once the selection
function of the survey is known, the correlation function is obtained
by simple pair countings even when the survey selection is
inhomogeneous and the geometrical shape of the survey is complex.
Although the theoretical calculation of the two-point correlation
function in redshift space is not as obvious as that of the power
spectrum, we now have an analytical formula which make a numerical
implementation very fast \citep{matsubara00,matsubara04}. The last
formula also contains wide-angle effects of the survey, which can not
be naturally implemented in the power spectrum analysis.

The purpose of this paper is to provide a thorough analysis of the
two-point correlation function in deep redshift space and its
dependence on cosmological parameters. The cosmological information
contained in the correlation function is quantitatively
investigated. As a result, we provide error forecasts of the
cosmological parameters probed by the correlation function in a given
redshift survey. Complementarity of samples at different redshifts is
quantitatively addressed. The analysis in this paper provides useful
information for the design of the future surveys at deep redshifts.

This paper is organized as follows. In \S 2, a compact version of the
analytic expression of the correlation function in arbitrarily deep
redshift space, derived from a more general expression of the previous
work is presented. Physical contents in the analytic formula are
addressed in \S 3. In \S 4, series of deep redshift surveys which are
within reach of the present-day technology are considered and expected
error bounds are estimated by means of the Fisher information matrix.
Optimal designs for the future redshift surveys are
indicated. Conclusions and discussion are given in \S 5. Some basic
notations are introduced in Appendix~\ref{app0}.  Useful equations in
calculating the Fisher matrix with marginalizations are given in
Appendix~\ref{app1}.

\section{The Analytic Expression of the Correlation Function in
Deep Redshift-space
\label{sec2}} 

We analyze the structure of the two-point correlation function in deep
redshift space, using the analytic formula derived in the most general
situation by relativistic linear perturbation theory
\citep{matsubara00}. In this section, the result of the analytic
formula is briefly reviewed. Some basic notations of the
Friedmann-Lema\^{\i}tre model with dark energy extension are introduced in
Appendix~\ref{app0}.

The general two-point correlation function in redshift space is a
function of the redshifts $z_1$, $z_2$ of the two points, and the
angle $\theta$ between them. The analytic result of
\citet{matsubara00} has the form,
\begin{eqnarray}
&&
  \xi(z_1,z_2,\theta) =
  D(z_1) D(z_2)
  \left[
    b(z_1) b(z_2) \xi^{(0)}(z_1,z_2,\theta)
  \right.
\nonumber\\
&&\quad
    +\,
    f(z_1) b(z_2) \xi^{(1)}(z_1,z_2,\theta) +
    f(z_2) b(z_1) \xi^{(1)}(z_2,z_1,\theta)
\nonumber\\
&&\quad
  \left. +\,
    f(z_1) f(z_2) \xi^{(2)}(z_1,z_2,\theta)
  \right],
\label{eq2-8}
\end{eqnarray}
where $b(z)$ is the redshift-dependent linear bias factor and $D(z)$
is the linear growth factor normalized as $D(z=0) = 1$, and calculated
by the differential equations (\ref{eq2-6a}) and (\ref{eq2-6b}). The
functions $\xi^{(n)}$ are defined below. In this paper, we assume the
concerning scales of the two-point correlation function are reasonably
smaller than the curvature scale of the universe: $x_{12} \ll
|K|^{-1/2}$, where $x_{12}$ is the comoving distance between the two
points. This exactly holds for a flat universe, and is fulfilled in
practice because the curvature of the universe is observationally at
least 20 times larger than the Hubble scale, $c/H_0$, since
$\mOmega_{K0} = \mOmega_{M0} + \mOmega_{\mLambda 0} - 1 = 0.02 \pm
0.02$ \citep{spergel03}. This approximation does not necessarily
require that the comoving distances to the two points from the
observer, $x(z_1)$ and $x(z_2)$, are smaller than the curvature
scale. In fact, the curvature effects are included in those distances
in the following compact version of the formula.

Adopting the above approximation, the components $\xi^{(k)}$ in the
analytic formula (\ref{eq2-8}) are given by
\begin{eqnarray}
&&
  \xi^{(0)}(z_1,z_2,\theta) = \xi^{(0)}_0(x_{12}),
\label{eq2-9a}\\
&&
  \xi^{(1)}(z_1,z_2,\theta) = 
  \frac13 \xi^{(0)}_0(x_{12}) + 
  A_{12}\cos\gamma_{12} \xi^{(1)}_1(x_{12})
\nonumber\\
&&\qquad\qquad\qquad\quad +\,
  \left(\cos^2\gamma_{12} - \frac13 \right) \xi^{(1)}_2(x_{12}),
\label{eq2-9b}\\
&&
  \xi^{(2)}(z_1,z_2,\theta) = 
  \frac{1}{15} \left(1 + 2\cos^2\theta\right)
  \xi^{(0)}_0(x_{12})
\nonumber\\
&&\qquad -\,
  \frac13 A_{12}A_{21} \cos\theta \xi^{(1)}_0(x_{12})
\nonumber\\
&&\qquad +\,
  \frac15
  \left[
    A_{12}
    \left(\cos\gamma_{12} - 2\cos\gamma_{21}\cos\theta\right)
  \right.
\nonumber\\
&&\qquad\qquad
  \left. +\,
    A_{21}
    \left(\cos\gamma_{21} - 2\cos\gamma_{12}\cos\theta\right)
  \right] \xi^{(1)}_1(x_{12})
\nonumber\\
&&\qquad -\,
  \frac17
  \biggl[
    \frac23 + \frac43\cos^2\theta -
    \left(\cos^2\gamma_{12} + \cos^2\gamma_{21}\right)
\nonumber\\
&&\qquad\qquad\qquad +\,
    4\cos\gamma_{12}\cos\gamma_{21}\cos\theta
  \biggr] \xi^{(1)}_2(x_{12})
\nonumber\\
&&\qquad +\,
  A_{12} A_{21}
   \left(\cos\gamma_{12}\cos\gamma_{21} + \frac13 \cos\theta\right)
   \xi^{(2)}_2(x_{12})
\nonumber\\
&&\qquad +\,
  \frac15
  \left[
    A_{12}
    \left(
      5\cos\gamma_{12}\cos^2\gamma_{21} - \cos\gamma_{12} +
      2\cos\gamma_{21}\cos\theta
    \right)
  \right.
\nonumber\\
&&\qquad\quad
   \left. + \,
      A_{21}
      \left(
         5\cos\gamma_{21}\cos^2\gamma_{12} - \cos\gamma_{21} +
         2\cos\gamma_{12}\cos\theta
      \right)
   \right]
\nonumber\\
&&\qquad\qquad\qquad\qquad\qquad
  \times\, \xi^{(2)}_3(x_{12})
\nonumber\\
&&\qquad +\,
   \frac17
   \biggl[
      \frac15 + \frac25\cos^2\theta -
      \left(\cos^2\gamma_{12} + \cos^2\gamma_{21}\right)
\nonumber\\
&&\qquad\qquad\qquad +\,
      4\cos\gamma_{12}\cos\gamma_{21}\cos\theta
\nonumber\\
&&\qquad\qquad\qquad  + \,
      7\cos^2\gamma_{12}\cos^2\gamma_{21}
   \biggr] \xi^{(2)}_4(x_{12}).
\label{eq2-9c}
\end{eqnarray}
In the above equations, the functions $\xi^{(n)}_l$ are defined by
\begin{equation}
  \xi^{(n)}_l(x) =
  \frac{(-1)^{n+l}}{x^{2n-l}}
  \int\frac{k^2dk}{2\pi^2}
  \frac{j_l(kx)}{k^{2n-l}} P(k),
\label{eq2-10}
\end{equation}
where $j_l$ is the spherical Bessel function, and $x_{12}$ denotes the
comoving distance between two points specified by variables $(z_1,
z_2, \theta)$. An explicit representation of $x_{12}$ is given by
\begin{eqnarray}
  {x_{12}} &\simeq& {S_K}(x_{12})
\nonumber\\
  &=&
  \left[
    {S}^2(z_1) + {S}^2(z_2) -
    2 C(z_1) C(z_2) S(z_1) S(z_2) \cos\theta
  \right.
\nonumber\\
&&\qquad\qquad
  \left. -\,
    K {S}^2(z_1) {S}^2(z_2) \left(1 + \cos^2\theta\right)
  \right]^{1/2},
\label{eq2-11}
\end{eqnarray}
where the first approximation is consisitent with $x_{12} \ll
|K|^{-1/2}$, and
\begin{eqnarray}
  S(z) &\equiv&
  S_K[x(z)]
\nonumber\\
  &=&
  \left\{
  \begin{array}{ll}
    (-K)^{-1/2} {\rm sinh}\left[(-K)^{1/2} x(z)\right], &
    (K < 0), \\
    x(z), & (K = 0), \\
    K^{-1/2} \sin\left[K^{1/2} x(z)\right], &
    (K > 0),
  \end{array}
  \right.
\label{eq2-12a}\\
   C(z) &\equiv&
   \frac{dS_K}{dx} (z) =
   \left\{
   \begin{array}{ll}
      \cosh\left[(-K)^{1/2} x(z)\right], &
      (K < 0), \\
      1, & (K = 0), \\
      \cos\left[K^{1/2} x(z)\right], &
      (K > 0).
   \end{array}
   \right.
\label{eq2-12b}
\end{eqnarray}
The angle $\gamma_{12}$ is the angle between the separation $x_{12}$
and the line of sight at the first point of $z_1$, and $\gamma_{21}$
is the similar angle at the second point of $z_2$. Trigonometric
geometry shows that they are analytically represented by
\begin{equation}
   \cos\gamma_{12} = 
   \frac{S(z_1) C(z_2) - C(z_1) S(z_2) \cos\theta}
     {S_K(x_{12})},
\label{eq2-13}
\end{equation}
and $\cos\gamma_{21}$ is given by a replacement of $z_1
\leftrightarrow z_2$ in the above equation. If the universe is exactly
flat, the above equation simply reduces to $\cos\gamma_{12} = (x_1 -
x_2\cos\theta)/x_{12}$. In this paper, we generally consider a
non-flat universe as well as a flat universe. The quantities $A_{12}$
and $A_{21}$ are needed when the redshift evolution of the clustering
between the two points and that of the selection function is not
negligible, and are given by
\begin{equation}
  A_{12} =
  S_K(x_{12}) H(z_1)
   \frac{d}{dz_1}\ln[H(z_1) D(z_1) f(z_1) n(z_1)],
\label{eq2-14}
\end{equation}
and the corresponding equation of $A_{21}$ with a replacement $z_1
\leftrightarrow z_2$ in this equation. In the above equation, $n(z) =
dN(<z)/dz$ is the selection function including the volume factor,
i.e., the differential number count of galaxies within a fixed angular
area as a function of redshift $z$. Since $H(z)d/dz = d/dx$, the
quantities $A_{12}$ and $A_{21}$ have the order of relative difference
of the product $H(z)D(z)f(z)n(z)$ between $z_1$ and $z_2$. Therefore,
if the difference of this product is negligible between two points at
$z_1$ and $z_2$, the terms with quantities $A_{12}$ and $A_{21}$ can
be omitted. Since the characteristic scales on which the factors
$H(z)$, $D(z)$, and $f(z)$ vary are of order of the Hubble scale, it
is unlikely that the contributions of these factors are important. In
practice, the determination of the galaxy correlation function on
Hubble scales is extremely difficult because of the high
signal-to-noise ratio. A possible dominant contribution to $A_{12}$
and $A_{21}$ is from the selection function $n(z)$, and therefore,
$A_{12} = S_K(x_{12}) H(z_1) d\ln[n(z_1)]/dz_1$, and $A_{21} =
S_K(x_{12}) H(z_2) d\ln[n(z_2)]/dz_2$ on sub-Hubble scales. If the
selection function varies on scales of interest, these terms should be
kept. When the selection function is approximately the same at
corresponding two points at $z_1$ and $z_2$, the terms with $A_{12}$
and $A_{21}$ can be omitted at all. This does not mean that these
terms can be omitted for completely homogeneous, volume-limited
sample, in which $n(z)$ increases with redshift. Instead, those terms
can be omitted when the spatial selection decreases with redshift and
cancels the increase of the volume factor per unit redshift in a
region of a fixed solid angle. In the following analysis of this
paper, these terms are assumed to be small and omitted for simplicity.
While the quantity $x_{12}$ is symmetric with the replacement of the
two points $1 \leftrightarrow 2$, the quantities $\gamma_{12}$ and
$A_{12}$ are not.

The quantity $\xi^{(0)}$ corresponds to the isotropic component of the
correlation function, since it depends only on $x_{12}$. The
quantities $\xi^{(1)}$ and $\xi^{(2)}$ are relevant to distortions by
the peculiar velocity field, since $f \rightarrow 0$ recovers the
correlation function in isotropic comoving space. The peculiar
velocity anisotropically distorts the correlation function even in
comoving space. The extended AP effect is included in the nonlinear,
anisotropic mapping from comoving space to redshift space (i.e.,
$z$-space) by $x_{12}(z_1,z_2,\theta)$ of equation (\ref{eq2-11}).

The above equations are complete set of equations we need to evaluate
the theoretical prediction of the correlation function in redshift
space when $x_{12} \ll |K|^{-1/2}$. A detailed derivation of the
equations and more general formula without the last restriction are
given in \citet{matsubara00}. As discussed in this reference, all the
known formula of the correlation function in redshift space are
limiting cases of the general formula presented here. When the galaxy
sample is sufficiently shallow, $z_1, z_2 \ll 1$, the formula above
reduces to the result derived by \citet{szalay98}, which includes
wide-angle effects on the peculiar-velocity distortions. When the
separation of the two-points is much smaller than the distances to
them, $x_{12} \ll x(z_1), x(z_2)$, the formula reduces to the result
by \citet{matsubara96} with the distant-observer approximation in deep
redshift space. When the above two limits are simultaneously applied,
the result of \citet{hamilton92} is recovered. By the Fourier
transform of the correlation function in distant-observer
approximation, the power spectrum in deep redshift space derived by
\citet{ballinger96} with distant-observer approximation is obtained.
The last expression with shallow limit $z \ll 1$ corresponds to the
original Kaiser's formula of the power spectrum in shallow redshift
space with distant-observer approximation \citep{kaiser87}.

%\begin{equation}
%\label{eq2-1}
%\end{equation}

\section{Cosmological Information in the Correlation Function in
Deep Redshift-space
\label{sec3}} 

The physical effects on the correlation function in deep
redshift-space can be placed in three categories. The first one
consists of the effects on the underlying mass power spectrum in
comoving space. The physical parameters which primarily determine the
shape of the power spectrum are $A_s$, $n_s$, $\mOmega_{\rm M0}h$,
$f_{\rm B}$, $h$, $\mOmega_{\nu 0}$, etc., where $A_s$ is the
amplitude of the power spectrum, $n_s$ is the primordial spectral
index, $h$ is the Hubble parameter normalized by $100$ km/s/Mpc,
$f_{\rm B} \equiv \mOmega_{\rm B0}/\mOmega_{\rm M0}$ is the fraction
of the baryon density to the total mass density, $\mOmega_{\nu 0}$ is
the density parameter of neutrinos.

The primordial spectral index $n_s$ determines the overall shape of
the power spectrum. In the standard Harrison-Zel'dovich spectrum,
$n_s= 1$. The parameter $\mOmega_{\rm M0}h$ determines the scale of
the particle horizon at the equality epoch $z_{\rm eq}$, since the
radiation density is accurately known by the temperature of the cosmic
microwave background, $T_0 = 2.725 \pm 0.001$ K \citep{mather99}. As a
result, the power spectrum has a characteristic peak at $k_{\rm eq}
\propto \mOmega_{\rm M0} h\ (\himpc)^{-1}$. The baryon fraction
$f_{\rm B}$ and Hubble parameter $h$, as well as $\mOmega_{\rm M0}h$
are responsible to the scales and strength of acoustic oscillations
and Silk damping. The dependences are not expressed by simple scaling
relations, and useful fitting relations on physical grounds are
provided by \citet{eisenstein98}. The density parameter of neutrinos,
$\mOmega_{\nu 0}$ characterizes the free streaming scales by the
existence of the hot dark matter. In standard cold dark matter
scenarios, this parameter is negligible. In the analysis of galaxy
clustering, the spectral amplitude $A_s$ is more conveniently
specified by the parameter $\sigma_8$. The relation between $A_s$ and
$\sigma_8$ depends on other parameters explained above which
determine the shape of the power spectrum.

\begin{figure*}
\epsscale{1.0} \plotone{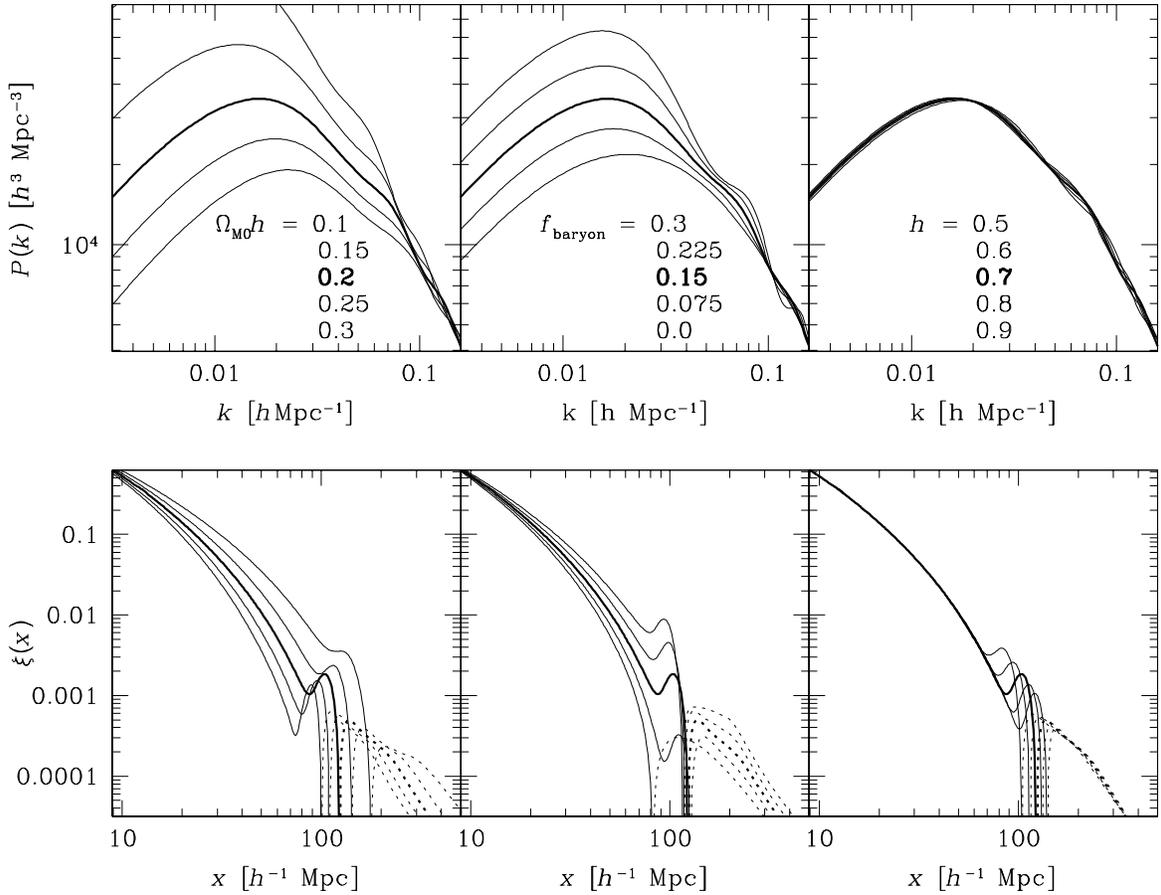} \figcaption[f1.eps]{ The power
spectrum (upper panels) and the correlation function (lower panels) in
comoving space. Thick lines correspond to model parameters
$\mOmega_{\rm M0} = 0.2$, $f_{\rm B} = 0.15$, and $h = 0.7$. The
amplitude is normalized by $\sigma_8 = 1$. Thin lines show the
variations of each parameters as suggested in the plots. When
$\mOmega_{\rm M0}h$ increases, the amplitude on large scales
decreases. When $f_{\rm B}$ increases, the amplitude on large scales
increases. When $h$ increases, the location of the baryon peak is
shifted to large scale.
\label{fig1}}
\end{figure*}

In Figure~\ref{fig1}, the dependences of the power spectrum and
correlation function in real space on parameters $\mOmega_{\rm M0}h$,
$f_{\rm B}$, and $h$ are plotted.
We consider a model with $n_s = 1$,
$\mOmega_{\rm M0} = 0.3$, $f_{\rm B} = 0.15$, $h = 0.7$, $\mOmega_{\nu
0} = 0$ as a fiducial case. Thick lines correspond to the fiducial
model, and other lines show the effects of varying individual
parameters, considering $\mOmega_{\rm M0}h$, $f_{\rm B}$ and $h$ as
independent parameters. The amplitude is normalized by $\sigma_8 = 1$.
The upper panels show the effects on the power spectrum and the lower
panels show that on the correlation function, which are the (3D)
Fourier transforms of the corresponding power spectrum. The
correlation function in comoving space is given by the function
$\xi^{(0)}_0 (x)$ of equation (\ref{eq2-10}).

The oscillatory behavior appears in the power spectrum by acoustic
waves before recombination epoch. Since this oscillatory behavior is
periodical in Fourier space, just one peak is appeared in the
correlation function. The scale of the peak corresponds to the sound
horizon at the recombination epoch. We call this peak as a {\em baryon
peak} in the correlation function.

Increasing $\mOmega_{\rm M0}h$ shifts the peak of the power spectrum
to the right, so that the powers on large scales are suppressed when
$\sigma_8$ is fixed. Correspondingly, the correlation function on
large scales are smaller for larger $\mOmega_{\rm M0}h$. The scale of
the zero-point of the correlation function decreases with this
parameter. The location of the baryon peak is changed by $\mOmega_{\rm
M0} h$ because the sound horizon is also dependent on this parameter.

The main effect of the parameter $f_{\rm B}$ is on the strength of the
Silk damping. Therefore, increasing $f_{\rm B}$ enhances the power on
large scales when $\sigma_8$ is fixed. The location of the baryon peak
is less dependent on $f_{\rm B}$, and the absolute amplitude of the
baryon peak is predominantly dependent on $f_{\rm B}$. The zero-point
of the correlation function is not much affected as long as $f_{\rm B}
\neq 0$.

Although the effect of $h$ on the power spectrum does not seem to be
significant in the Figure, the phase of the baryon oscillation is
shifted. This is partly because we fix $\mOmega_{\rm M0}h$ instead of
the physical dark matter density $\mOmega_{\rm M0}h^2$. Since the
fraction $f_{\rm B}$ is also fixed, the physical densities varies with
$h$ as $\mOmega_{\rm M0}h^2 \propto h$, $\mOmega_{\rm B0}h^2 \propto
h$ in our choice of the independent parameters. Although the shape of
the power spectrum is determined by the physical density parameters,
the length scale should be measured in units of $\himpc$ in redshift
space, and one can not eliminate the explicit dependence on the Hubble
parameter by scalings. It is only when baryons are absent that the
shape of the power spectrum is characterized by a single parameter
$\mOmega_{\rm M0}h$, without any explicit dependence on $h$.

The effect of $h$ on the correlation function is more noticeable since
the multiple baryon wiggles in the power spectrum cumulatively
contribute to the baryon peak in the correlation function. The sound
horizon at the recombination epoch increases with $h$ when
$\mOmega_{\rm M0}h$ and $f_{\rm B}$ are fixed, and the location of the
baryon peak is shifted to the right.

\begin{figure}
\epsscale{1.0} \plotone{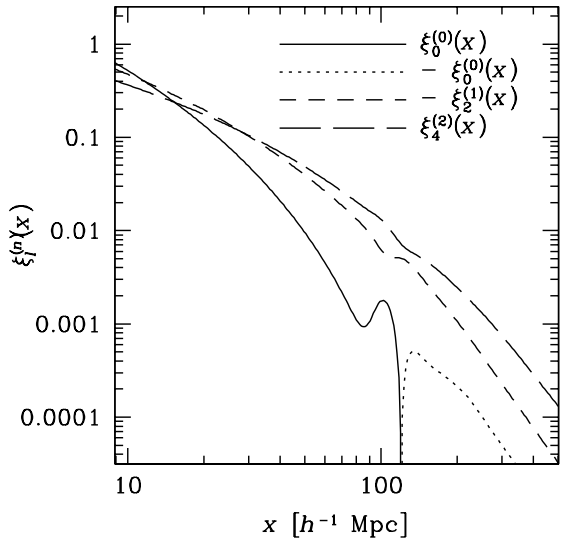} \figcaption[f2.eps]{ Examples of the
function $\xi^{(n)}_l(x)$ that is needed in the analytic formula of
the correlation function in redshift space.
\label{fig2}}
\end{figure}

The second category of the physical effects on the correlation
function in deep redshift-space is the velocity distortions, or the
Kaiser's effects. These effects are included in second and third terms
of equation (\ref{eq2-8}). In linear regime, the coherent infall
toward centers of density maxima flatten the clustering pattern along
the lines-of-sight. Depending on the geometrical angles among the
observer and the two points, the correlation function in redshift
space is a linear combination of the functions $\xi^{(n)}_l(x_{12})$.
The functions $\xi^{(0)}_0$, $\xi^{(1)}_2$, and $\xi^{(2)}_4$, are
plotted in Figure~\ref{fig2} with the fiducial model of the power
spectrum described above.
Other functions, $\xi^{(1)}_0$,
$\xi^{(1)}_1$, $\xi^{(2)}_2$, and $\xi^{(1)}_3$ are not significant
when $A_{12}$, $A_{21}$ are small and are omitted in the Figure. All
the functions $\xi^{(n)}_l$ have irregularities on the scale of the
baryon peak. Since the correlation function in redshift space is
expressed by a linear combination of these functions, the resulting
function has the same irregularities on the same scale.

The third category of the physical effects on the correlation function
in deep redshift space is the extended AP effect. The irregularities
on the scale of baryon peak play a role of spherical objects with
known radius in comoving space. Originally, \citet{alcock79} proposed
that measuring the ellipticity of some objects which are spherical in
comoving space, the value of the cosmological constant can be
constrained. It was pointed out that the power spectrum
\citep{ballinger96} and the correlation function \citep{matsubara96}
can serve as such objects. The baryon wiggles in the power spectrum
can be used for this purpose \citep{blake03,seo03}. While the baryonic
feature is split into many wiggles in the power spectrum, there is
just one peak in the correlation function.

\begin{figure*}
\epsscale{1.0} \plotone{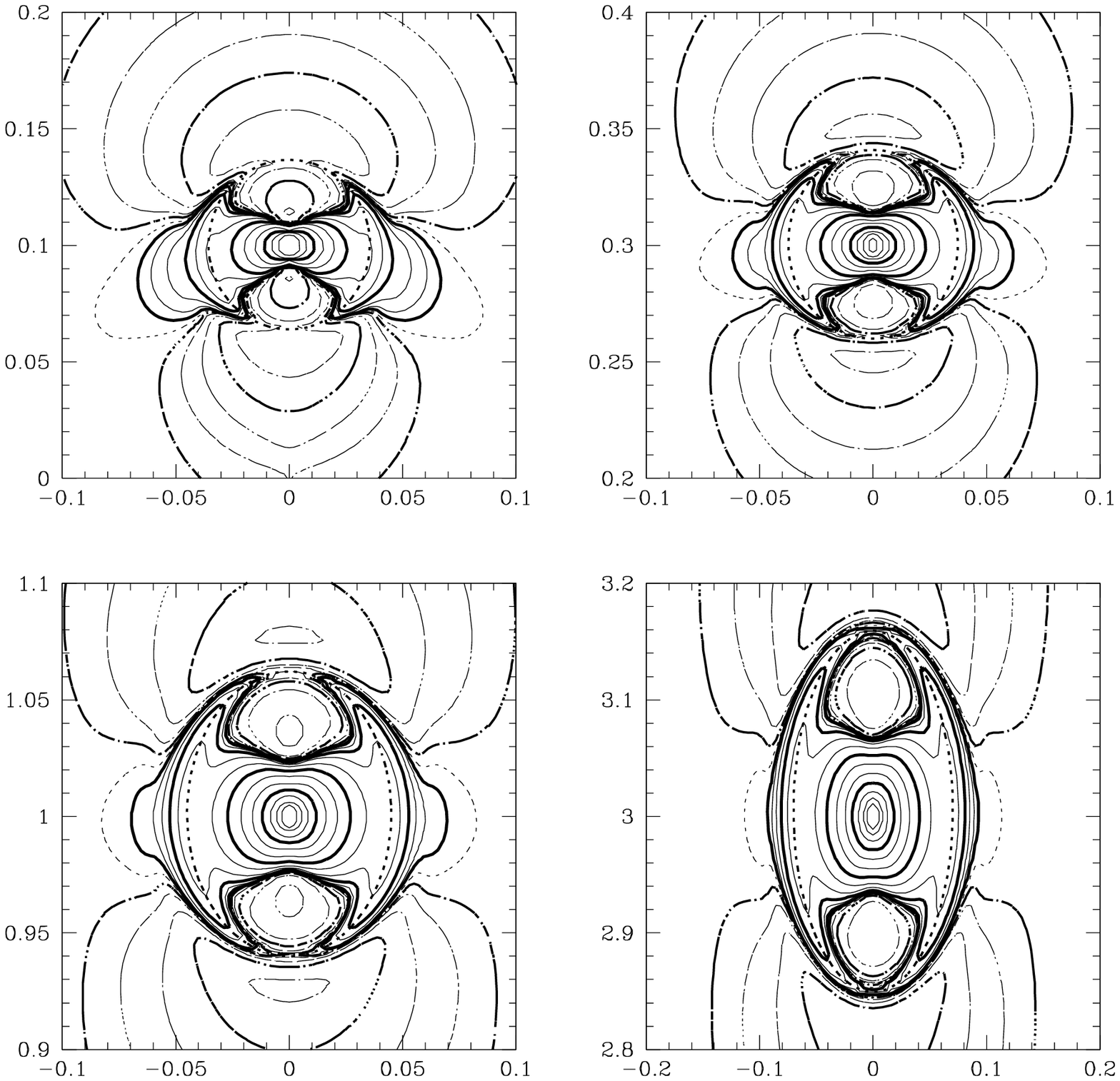} \figcaption[f3.eps]{ Contour plots of
the correlation function $\xi(z_1,z_2,\theta)$ in deep redshift space.
The first point with redshift $z_1$ is sitting at the center of the
plot. The contours represent the correlation value as a function of
$z_2$ and $\theta$ and the coordinates are defined by $(z_2\sin\theta,
z_2\cos\theta)$. In other words, the plots show the contours of the
correlation function in apparent $z$-space. The fiducial model defined
in the text is assumed.
\label{fig3}}
\end{figure*}

In Figure~\ref{fig3}, contour plots of the resulting correlation
function in redshift space with $z_1 = 0.1$, $0.3$, $1.0$, and $3.0$
are shown.
The fiducial model parameters with $\mOmega_{\rm K0} = 0$,
$w=-1$, $\mOmega_{\rm M0} = 0.7$, $n_s = 1$, $\mOmega_{\nu 0} = 0$,
$\mOmega_{\rm M0} h = 0.21$, $f_{\rm B} = 0.15$ are adopted. The bias
parameters are assumed to be $b(z) =1$, $2$, $3$, and $3$ for $z =
0.1$, $0.3$, $1.0$, and $3.0$, respectively. The normalization of the
power spectrum is chosen so that the galaxy normalizations at
corresponding redshifts are unity; $\sigma_{8g}(z) \equiv
D(z)b(z)\sigma_8 = 1$. First point labeled by $z_1$ is sitting at the
center of each plot. The contour lines represent the value of the
correlation function in redshift space, $\xi(z_1,z_2,\theta)$, and the
coordinates of the plots are $(z_2 \sin\theta, z_2 \cos\theta)$. In
other words, the plots show the contours of the correlation function
in apparent $z$-space. A prominent feature in these plots is the
existence of ridges. These ridges are perfectly circular in comoving
space, i.e., they are located at the lines of constant $x_{12}$. The
scales of the baryon peak in comoving correlation function, which we
have seen in the Figure~\ref{fig1}, is indicated by thick dotted lines
in the plots. Obviously, the ridges correspond to the baryon peaks in
comoving correlation function. Therefore, we will call those ridges as
the {\em baryon ridges}.  While the peculiar velocities alter the
amplitudes of the correlation function along the ridges depending on
the direction relative to the lines of sight, the shape of the ridges
are not distorted by them.  Therefore, the baryon ridges are ideal
statistically spherical objects which are useful for the extended AP
test. The Fourier counterpart of the baryon ridge is the acoustic
rings in the power spectrum in redshift space \citep{hu03}. While the
rings in the power spectrum are spread over many scales, the baryon
ridges in the correlation function have just a single scale.

\begin{figure*}
\epsscale{0.7} \plotone{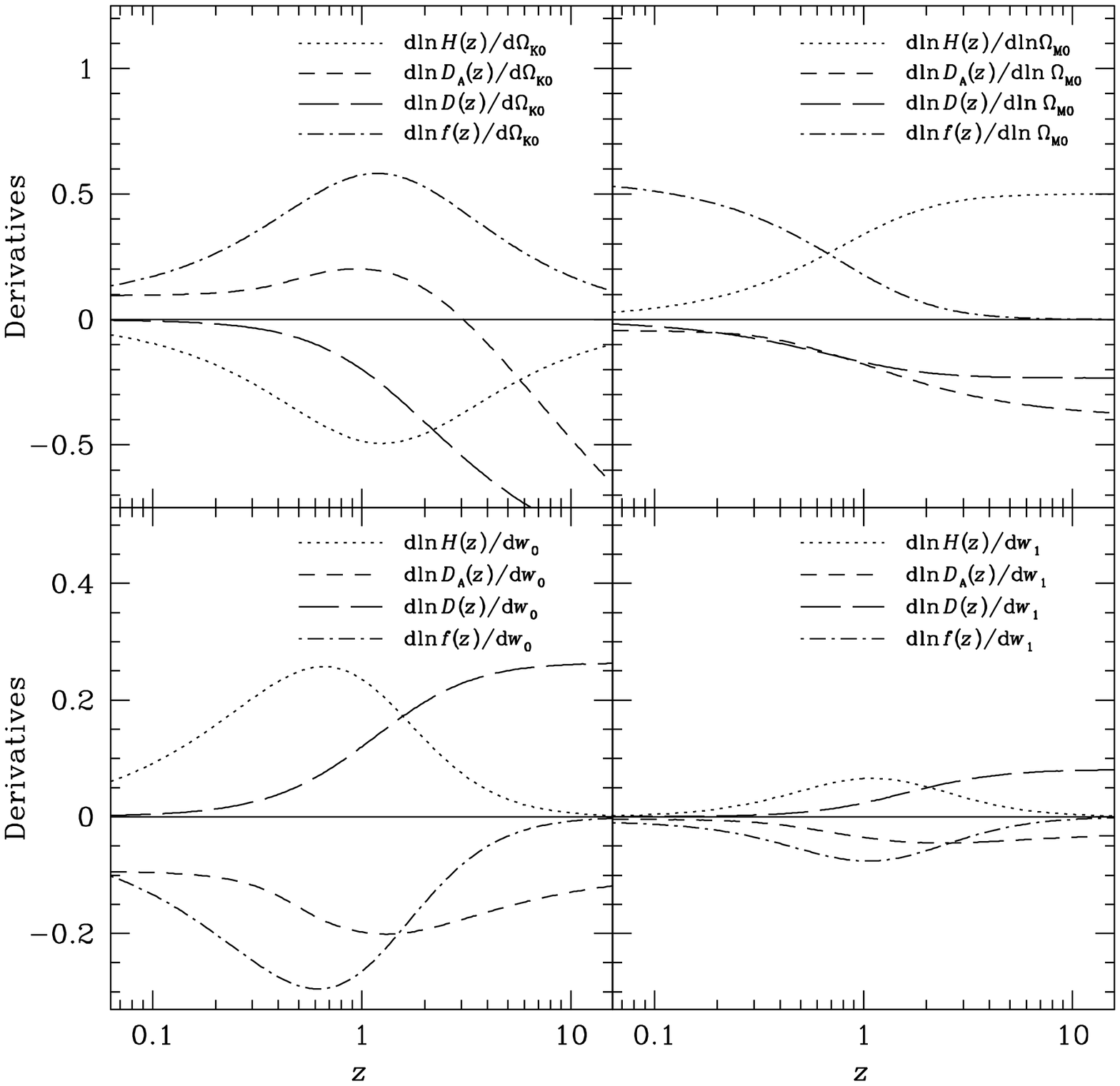} \figcaption[f4.eps]{ Sensitivities of
the cosmological functions on parameters. To present the fractional
sensitivities, derivatives of the logarithm of the Hubble parameter
$H(z)$ (dotted lines), the comoving angular diameter distance $D_{\rm
A}(z)$ (short dashed lines), the linear growth rate $D(z)$ (long
dashed lines), and logarithmic growth rate $f(z)$ (dot-dashed lines)
are shown as functions of the redshift. Upper left panel shows the
sensitivities on the curvature parameter $\mOmega_{\rm K0}$, fixing
$\mOmega_{\rm M0}$. Upper right panel on the density parameter
$\mOmega_{\rm M0}$, fixing $\mOmega_{\rm K0}$. Lower panels on the
parameters of equation of state of the dark energy, $w_0$ and $w_1$.
\label{fig4}}
\end{figure*}

Besides the linear power spectrum at present, the correlation function
in deep redshift space depends on other cosmological parameters such
as $\mOmega_{\rm K0}$, $\mOmega_{\rm M0}$, and $w$. The overall
amplitude of the correlation function depends on the linear growth
rate $D(z)$. The peculiar velocity effects depend on the logarithmic
growth rate $f(z)$. The AP effects depends on the time-dependent
Hubble parameter $H(z)$ and comoving angular diameter distance,
$D_{\rm A}(z) \equiv S(z)$. The function $H(z)$ specifies the scales
along the lines of sight in redshift space, and the function $D_{\rm
A}(z)$ determines the scales perpendicular to the lines of sight in
redshift space. Figure~\ref{fig4} shows the dependences of the
relevant functions on cosmological parameters.
In the derivatives by
$\mOmega_{\rm K0}$, the mass density parameter $\mOmega_{\rm M0}$ is
fixed so that the derivatives with respect to $\mOmega_{\rm K0}$ is
equivalent to the derivatives with respect to the dark energy density
parameter, $\mOmega_{\rm Q0}$. The redshift evolution of dark energy
equation of state is parameterized by $w = w_0 + w_1 z/(1+z)$. The
parameter sensitivity of the physical quantities are generically
higher in deep redshift samples. The dark energy is less important
when the redshift is too high. Only the function $f(z)$ is sensitive
to $\mOmega_{\rm M0}$ at low redshifts. Overall, parameter sensitivity
is high at around $z \sim 1.0$. Therefore, intense redshift surveys
around redshift $\sim 1$ provide important constraints on cosmological
models.

%\begin{equation}
%\label{eq2-1}
%\end{equation}

\section{Constraining Cosmological Parameters with Deep Redshift Samples
\label{sec4}}

\subsection{Evaluating the Fisher Information Matrix
\label{sec4-1}}

Accurate estimations of how the correlation function in deep redshift
space can constrain the cosmological parameters are crucial in
designing redshift surveys. For this purpose, we assume idealized
surveys which are accessible by present-day technologies, and give
estimates of the error bounds for cosmological parameters, calculating
Fisher information matrices. In this section, we consider
$\mOmega_{\rm K0}$, $\mOmega_{\rm M0}$, $\mOmega_{\rm M0} h$, $f_{\rm
B}$, $w$, and $b$ as an independent set of cosmological parameters.
The bias is the most uncertain factor in galaxy redshift surveys. On
linear scales, the complex bias uncertainty is renormalized to a
scale-independent linear bias parameter, $b$
\citep{scherrer98,matsubara99}. The bias parameter is not a universal
parameter and varies from sample to sample, depending on selection
criteria of individual observations and also on redshifts. In the
following, we always consider the situation that the bias parameter in
a given sample is unknown in each sample and should be determined
simultaneously with other parameters. In other words, the bias is
always marginalized over in the Fisher analysis below.  Therefore, the
results in the following are free from the bias uncertainty unless the
variation of the bias in each sample is too strong.

The forecasts of expected errors in parameter determinations in a
given sample are most easily obtained by the analysis of the Fisher
information matrix \citep[e.g.,][]{kendall69,therrien92}. This
technique enables us to obtain expected errors in parameter
determination by a given sample without performing any Monte-Carlo
simulation. The Fisher information matrix is the expected curvature
matrix around a maximum point of the logarithmic likelihood function
${\cal L}$ in parameter space as defined by the equation
(\ref{eqa-1}). When the distribution of the data is given
approximately by a multivariate Gaussian one with a correlation matrix
$\bfm{C}$, the Fisher matrix is given by
\citep[e.g.,][]{vogeley96,tegmark97}
\begin{equation}
  F_{\alpha\beta} =
  \frac12 {\rm Tr}
  \left(
     \bfm{C}^{-1} \frac{\partial\bfm{C}}{\partial\theta_\alpha}
     \bfm{C}^{-1} \frac{\partial\bfm{C}}{\partial\theta_\beta}
  \right),
\label{eq4-1}
\end{equation}
where $\{\theta_\alpha\}$ is the set of model parameters. Predictions
of the error bounds when some parameters are marginalized over are
also useful. For this purpose, we can define the marginalized Fisher
matrix by a method presented in Appendix~\ref{app1}, and obtain the
corresponding error forecasts. The bias parameters are always
marginalized over by that method.

Thus, the calculation of the Fisher matrix is straightforward once the
correlation matrix $\bfm{C}$ of the data and its derivatives are
obtained. However, the calculation of the correlation matrix depends
on what kind of observed data is analyzed. A straightforward
definition of the data in an analysis of the two-point correlation
function is the binned values of the correlation function. The
two-point correlation function in redshift space is a function of a
set of the variables $(z_1, z_2, \theta)$. With a suitable binning of
this three-dimensional space, we obtain the data vector $\bfm{\xi}$
(assumed to be a column vector), which consists of the values of the
correlation function averaged over within each bin. In this case, the
correlation matrix is given by $\bfm{C} =
\langle\bfm{\xi}\bfm{\xi}^{\rm T}\rangle$. To evaluate this
correlation matrix, appropriate estimates of covariances between
components of $\bfm{\xi}$ in a given observation with certain geometry
and selection function are necessary. This is marginally possible, but
requires intensive multidimensional integrations to obtain the proper
correlation matrix.

Fortunately, there is a simpler, alternative method of evaluating the
Fisher matrix of a given sample, using the information of the
two-point correlation function \citep{matsubara02,matsubara03}. In
this method, the data vector is taken to be pixelized galaxy counts
$N_i$ in a survey sample. Thus, the correlation matrix is simply given
by the smoothed correlation function convolved by pixels:
\begin{eqnarray}
  C_{ij} &=&
  \left\langle
    \left(N_i - \bar{N}_i\right)
    \left(N_j - \bar{N}_j\right)
  \right\rangle
\nonumber\\
  &=&
  \frac{\bar{N}_i \bar{N}_j}{v_i v_j}
  \int_{v_i} d^3 z_1
  \int_{v_j} d^3 z_2\ 
  \xi(z_1,z_2,\theta_{12}) +
  \bar{N}_i \delta_{ij},
\label{eq4-2}
\end{eqnarray}
where $v_i$ is the volume of the pixel $i$ and $\bar{N}_i$ is the
expected number of galaxies in that pixel. The integrated regions are
limited within pixels in redshift space. The second term is the
contribution from the shot noise \citep{peebles80}. At first sight,
the six-dimensional integration in equation (\ref{eq4-2}) seems
computationally costly to perform. Nevertheless, \citet{matsubara04}
showed that there is no need to perform these direct integrations when
one can set the spherical shape of the pixels in comoving space. In
which case, performing the smoothing integrations of equation
(\ref{eq4-2}) is equivalent to just replacing the power spectrum
$P(k)$ by $P(k)W^2(kR)$ in the equation (\ref{eq2-10}), where $W(x) =
(\sin x - x \cos x)/3x^3$ is the smoothing kernel in Fourier space and
$R$ is the smoothing radius in comoving space. On one hand, one need
to know the cosmological parameters $\mOmega_{M0}$, $\mOmega_{Q0}$,
and $w$ in advance to set pixels in redshift space which are exactly
spherical in comoving space. On the other hand, the slight ellipticity
of the pixels does not change the value of the correlation matrix
which is calculated by assuming spherical pixels. Therefore, with an
approximate set of cosmological parameters, one need not to perform
the integrations of equation (\ref{eq4-2}) and directly obtain an
accurate correlation matrix in just the same way as computing the
correlation function itself in redshift space.

In the following, we use the latter method to calculate the Fisher
matrix. The correlation matrix of equation (\ref{eq4-2}) is calculated
from the expected number density in each cells and models of the
correlation function. The Fisher matrix with any marginalization is
obtained only from the correlation matrix.

\subsection{Baseline Redshift Surveys
\label{sec4-2}}

One of the advantages of the deep redshift surveys is that the
information from the time sequence of the galaxy clustering is
available. Since the solid angle on the sky is limited to $4\pi$ at
most, the comoving volume in which we can observe galaxies is larger
in high-redshift universe than in low-redshift universe. If the
galaxies in a spectroscopic survey are selected in a fixed solid angle
by certain criteria, as commonly done in redshift surveys like the 2dF
and the SDSS surveys, the number density of observed galaxies becomes
sparse at high redshifts. This is a critical drawback in a correlation
analysis since the shot noise dominates the clustering signal on
scales of interest. For example, SDSS quasar sample is not optimal to
determine the cosmological parameters from the clustering, because of
their sparseness \citep{matsubara02}. To maximally take advantage of
the linear formula of the correlation function in deep redshift space,
clustering properties on scales of $10$--$200\himpc$ provides a useful
information on cosmology. Too small scales suffer the nonlinear
effects which has less cosmological information and have difficulties
in analytical treatments. Therefore, observations with a number
density of over about $1/(10\himpc)^3$ is ideal.

\begin{figure}
\epsscale{1.1} \plotone{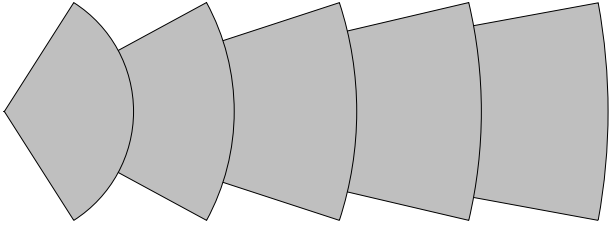} \figcaption[f5.eps]{ Illustration of
an optimal survey design with multi-layers of regions.
\label{fig5}}
\end{figure}

To retain comparable number densities at different redshifts, the
survey region on the sky should be narrower at high redshifts than at
low redshifts, since the total number of observed galaxies is usually
limited by the observation time, or budgets. Therefore, an optimal
design of the redshift survey will be something similar to that
illustrated in Figure~\ref{fig5}.
The optimal survey illustrated here
consists of several layers of sub-surveys. Each sub-survey is
optimized to catalog particular redshift range and has selection
criteria that is most effective to pick the galaxies up in
corresponding range of redshift. While simple color selections can be
used for this purpose, the photometric redshift data might be the best
to fulfill the required selections. In low-redshift layers, the
surface number density of galaxies for the spectroscopy can be low,
while the sky coverage of the survey regions should be wide. In
high-redshift layers, the surface number density should be high, while
the sky coverage can be narrow. The requirements of the telescope for
each layers are different. Thus this optimal survey would effectively
carried out by several telescopes with a wide spectrum of capabilities
such as resolutions and wideness of the field of view.

\begin{figure}
\epsscale{1.1} \plotone{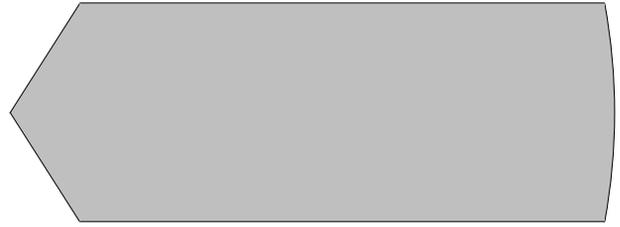} \figcaption[f6.eps]{ Another survey
design with continuous layers.
\label{fig6}}
\end{figure}

One can also imagine a limit of infinite number of layers. The survey
geometry in this case is like a big column in redshift space, as
illustrated in Figure~\ref{fig6}.
In this case, the selection criteria
should continuously vary with the position on the sky, therefore
delicate selections are required to sample galaxies as homogeneously
as possible. While technically more challenging, this survey strategy
is ideal for the cosmological analysis of the clustering at various
redshifts in the deep universe.

In this section, however, we consider more realistic surveys that can
be achieved by present-day technologies. We take the series of
baseline redshift surveys explained below. The surveys are categorized
to four types, which are assumed to be volume-limited samples. The
choices of samples at high-redshifts are similar to that considered by
\citet{seo03}.

The first sample is a low-redshift sample around $z\sim 0.15$. This
sample is motivated by a volume-limited subsample of galaxies in the
SDSS survey \citep{tegmark04} with absolute $r$-magnitude $M_r \sim
-22$. Our baseline sample has a uniform number density $\bar{n} = 1
\times 10^{-3} /(\himpc)^3$ in the redshift range $z=0.05$--$0.25$.
The sky coverage is assumed to be $\mOmega_{\rm sky} = 8.05$ str. The
bias factor is assumed to be $b=1.3$. We call this sample as {\em
Sample A} below.

The second sample is the one around $z\sim 0.3$. This sample is
motivated by the volume-limited catalog of the Luminous Red Galaxies
(LRGs) in the SDSS survey \citep{eisenstein01}. Our sample is assumed
to have $\bar{n} = 1\times 10^{-4} /(\himpc)^3$, $z=0.2$--$0.4$,
$\mOmega_{\rm sky} = 2.57$ str., and $b=2.0$. We call this sample as
{\em Sample B} below.

The next samples are the series of surveys around $z\sim 1$. The
choice of the target galaxies is not trivial. We follow the baseline
surveys that are considered by \citet{seo03}. The target galaxies are
assumed to be either giant ellipticals or luminous star-forming
galaxies. This category of samples consists of surveys of four
redshift bins, $z=0.5$--$0.7$, $z=0.7$--$0.9$, $z=0.9$--$1.1$, and
$z=1.1$--$1.3$. The sky coverage is assumed to be $\mOmega_{\rm sky} =
0.914$, $0.647$, $0.517$, and $0.451$, respectively. The bias factors
are $1.25$, $1.40$, $1.55$, and $1.70$, respectively. Approximately we
have $\sigma_{\rm 8g} \simeq 1$ in each samples with the above choices
of bias factors. We assume $\bar{n} = 5\times 10^{-4} /(\himpc)^3$
commonly for each samples. We call the series of samples as {\em Sample
C1} to {\em Sample C4} below.

The last sample is the survey of Lyman break galaxies around $z\sim
3$. The selection techniques for this kind of galaxies are reported by
\citet{steidel96}. We consider the sample with a redshift range of
$z=2.65$--$3.35$. We assume $b=3.3$ and $\bar{n} = 1\times 10^{-3}
/(\himpc)^3$ which are consistent with the observation
\citep{steidel98,adelberger98}. We call this sample as {\em Sample D}.
The parameters of the baseline surveys are summarized in
Table~\ref{tab1}.
\begin{table}
\begin{center}
\caption{Parameters of the baseline surveys.
\label{tab1}}
\begin{tabular}{c|ccccccc}
\tableline\tableline
Samples & $z$ & area [str.] & $b$ 
& $\bar{n}$ [$/(10\himpc)^3$] & Candidates\\
\tableline
A  & 0.05--0.25 & 8.05 & 1.30  & $1.0$ & $M_r \sim -22$ galaxies \\ \hline
B  & 0.2--0.4   & 2.57 & 2.00  & $0.1$ & Luminous red galaxies \\ \hline
C1 & 0.5--0.7   & 0.914 & 1.25 & $0.5$ & \\
C2 & 0.7--0.9   & 0.647 & 1.40 & $0.5$ & Giant ellipticals, or\\
C3 & 0.9--1.1   & 0.517 & 1.55 & $0.5$ & star-forming galaxies\\
C4 & 1.1--1.3   & 0.451 & 1.70 & $0.5$ & \\ \hline
D  & 2.65--3.35 & 0.102 & 3.30 & $1.0$ & Lyman break galaxies\\
\tableline
\end{tabular}
\end{center}
\end{table}

The Fisher matrix is calculated by assuming a cone geometry with an
open angle which gives 1/10 of the assumed sky coverage in each
sample. The spherical cells with a comoving radius $15\himpc$ are
placed in closed-packed structure in a survey region. The total number
of cells is exactly 4900 in each 1/10 subsample. Thus the comoving
volumes of all samples are essentially the same. The resulting Fisher
matrix is multiplied by 10 to obtain the final Fisher matrix. This
method of subdividing the sample into 1/10 is taken to reduce the
dimension of the matrix and to compromise with the CPU time and the
memory requirement of handling the large correlation matrix. In
general, the subdivision overestimates the error bounds because the
information from cross correlations between subsamples is neglected.
However, the larger the sample is, the less such cross correlations
are. We confirmed that the error bounds of the full sample is
accurately calculated by just adding the Fisher matrices of the
subsamples in our case.

\subsection{Results of the Fisher Analysis}

According to the calculation of the Fisher matrix, the expected error
bounds with any marginalization are straightforwardly obtained by
methods explained in Appendix~\ref{app1}. Since the bias parameter
should be simultaneously determined sample by sample, we calculate the
Fisher matrix of each sample with bias separately marginalized over
within each sample. Therefore, the results below are not affected by
bias uncertainties, as long as the bias factor does not significantly
varies within each samples.

\begin{figure*}
\epsscale{1.1} \plotone{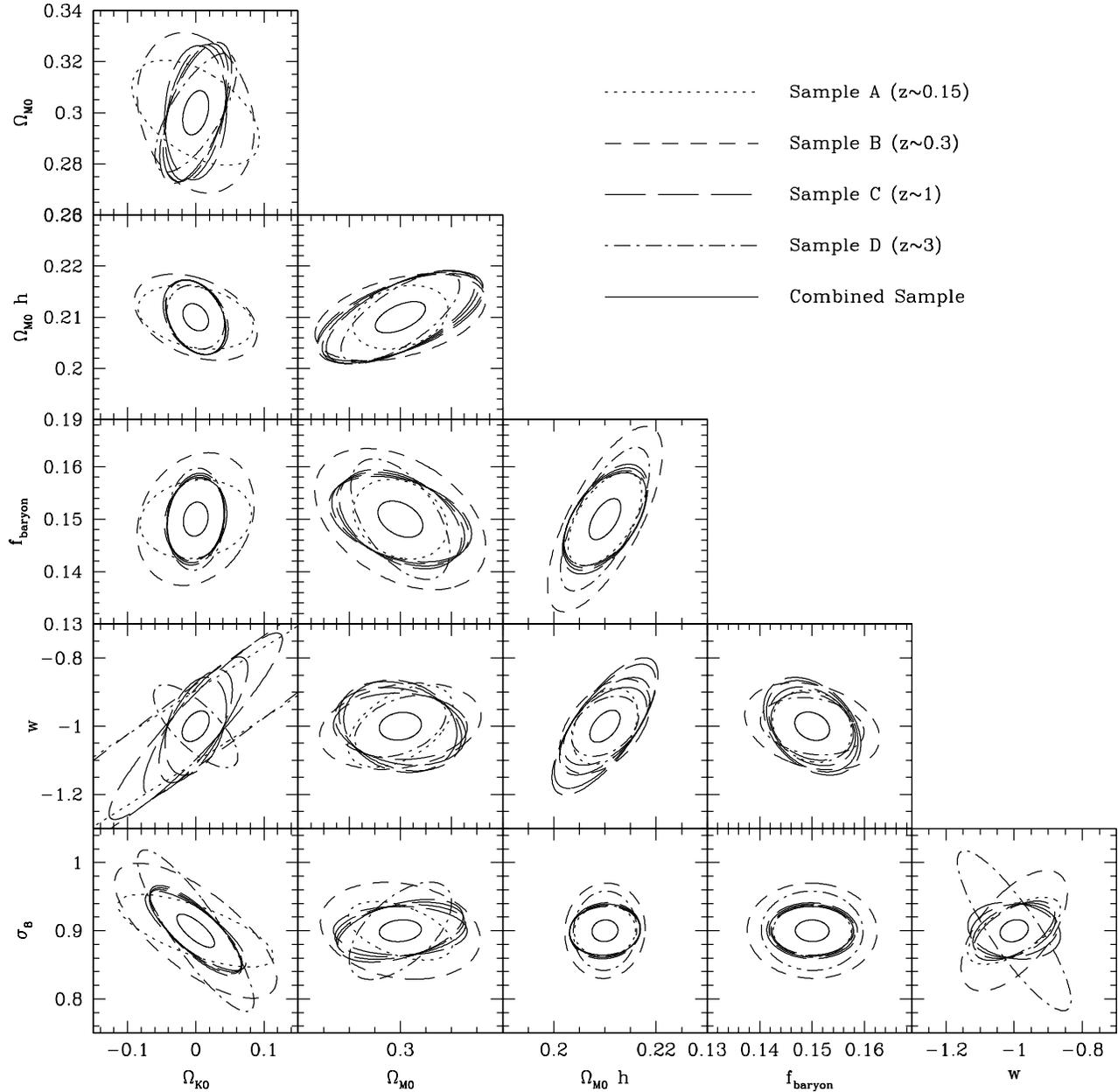} \figcaption[f7.eps]{ Two-parameter
joint error bounds expected by the baseline surveys. Each ellipse
represents the expected constraint on the parameter space with
$1\sigma$ significance. In each panel, corresponding two parameters
are varied with bias marginalized over. Other parameters are fixed in
each plot. Error bounds for Sample A (dotted lines), Sample B
(short-dashed lines), Sample C1--C4 (long-dashed lines), Sample D
(dot-dashed lines) are plotted together with a combined sample (solid
lines).
\label{fig7}}
\end{figure*}

In Figure~\ref{fig7}, concentration ellipsoids in 2-parameter
estimations are plotted, i.e., each panel shows the expected error
bounds when corresponding two parameters are simultaneously
determined, fixing other parameters except the bias parameter.
The bias parameters are marginalized over. The Fisher matrix of the
combined sample are also calculated by adding the Fisher matrices of
Sample A--D with bias separately marginalized over sample by sample.

Most of the pairs of parameters do not exhibit degeneracy between
them. This means that those pairs of parameters contribute to the
correlation function fairly differently as explained in \S 3. However,
one can notice relatively clear degeneracies between $\mOmega_{\rm
K0}$ and $w$. Since we consider $\mOmega_{\rm M0}$ and $\mOmega_{\rm
K0}$ as independent parameters, changing the parameter $\mOmega_{\rm
K0}$ is equivalent to changing the dark energy density $\mOmega_{\rm
Q0}$ with $\mOmega_{\rm M0}$ fixed. The parameters of the dark energy
are mainly constrained by AP effect besides the linear growth rate. In
the low-redshift samples, the parameters $\mOmega_{\rm Q0}$ and $w$
only depend on the growth rate and thus degenerate with each other.
Since the bias parameter and the growth rate is distinguishable by the
velocity distortion, one can still extract information on the dark
energy from low-redshift sample with the cost of the degeneracy. On
the other hand, high-redshift samples can actually constrain these two
dark energy parameter independently.

\begin{figure}
\epsscale{1.1} \plotone{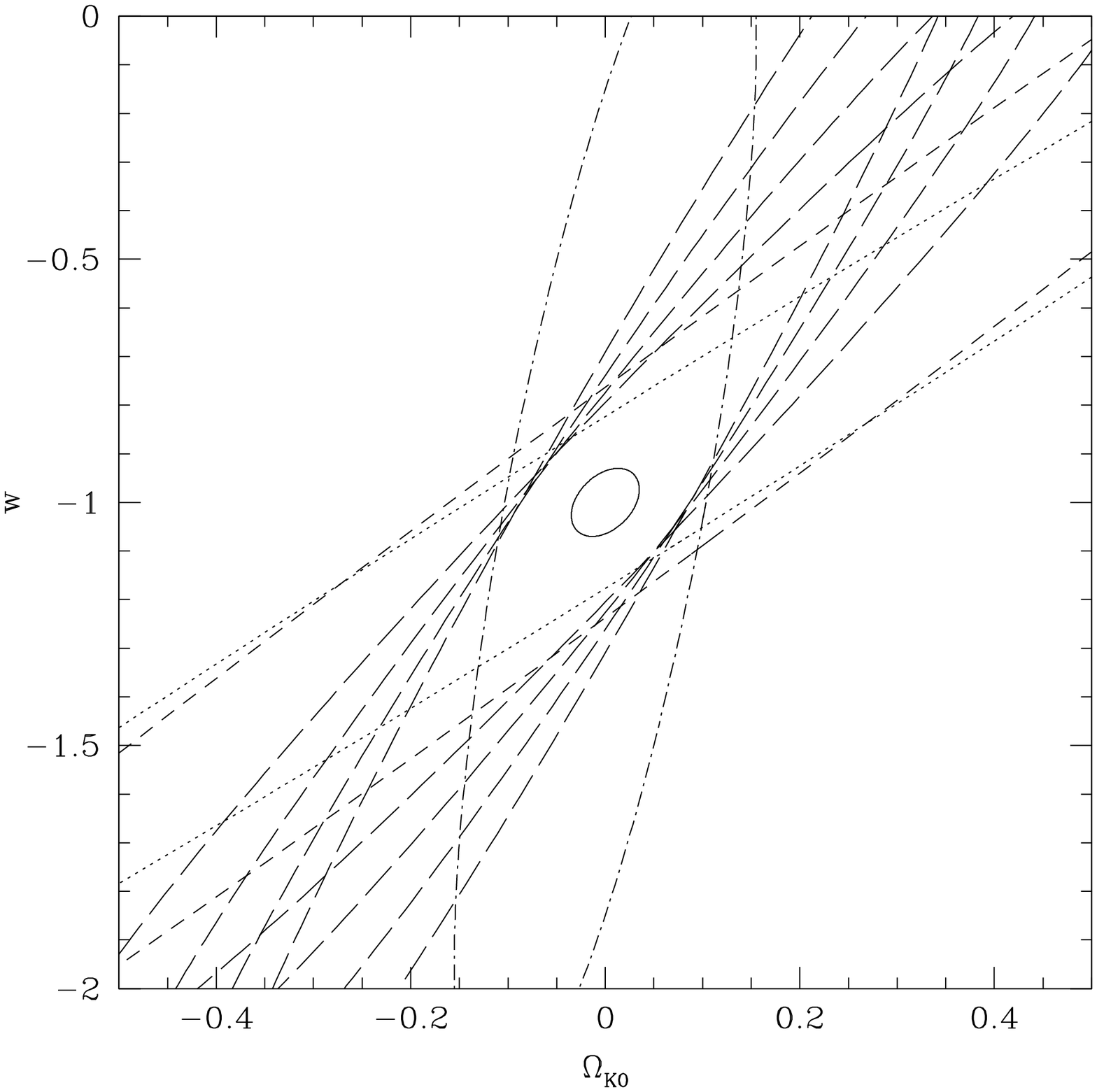} \figcaption[f8.eps]{ Joint error
bounds of the dark energy parameters expected by the baseline surveys.
Parameters $\mOmega_{\rm M0}$, $\mOmega_{\rm M0} h$, $f_{\rm B}$,
$\sigma_8$ and $b$ are all marginalized over. The correspondences
between lines and samples are the same as in Figure~\ref{fig7}.
\label{fig8}}
\end{figure}

The direction of the major axis of the concentration ellipse in the
$\mOmega_{\rm K0}$--$w$ plane rotates anti-clockwise with the average
redshift of the samples. Accordingly, the concentration ellipse of the
combined sample is quite small. For example, the error bound of
$\mOmega_{\rm Q0}$ is $\sim \pm 2\%$ and that of $w$ is $\sim \pm
5\%$. The presented error bounds are given when the other parameters
are fixed except the bias. It is also interesting to see the expected
constraints when all parameters are determined by only using the deep
galaxy surveys. In Figure~\ref{fig8}, the concentration ellipses are
plotted on $\mOmega_{\rm K0}$--$w$ plane when the other parameters
$\mOmega_{\rm M0}$, $\mOmega_{\rm M0} h$, $f_{\rm B}$, $\sigma_8$ and
$b$'s are all marginalized over.
The complementarity of the surveys at various redshifts is obvious in
this case. The concentration ellipses of high-redshift samples are
also elongated by uncertainties of other parameters. The direction of
the major axis of the ellipses rotates anti-clockwise with redshifts.
Consequently, the degeneracy between $\mOmega_{\rm K0}$ and $w$ are
broken in the combined sample and the error bound is much smaller than
in the individual samples. The error bound of $\mOmega_{\rm Q0}$ is
$\sim \pm 4\%$ and that of $w$ is $\sim \pm 10\%$ even when all the
other parameters are completely unknown.

%\begin{equation}
%\label{eq2-1}
%\end{equation}

\section{Conclusions and Discussion
\label{sec5}}

In this paper, properties of the two-point correlation function in
deep redshift space is theoretically investigated. A completely
general expression of the linear correlation function derived by
\citet{matsubara00} is reduced to a simpler form,
eq.(\ref{eq2-8})--(\ref{eq2-14}) with an approximation which is valid
for realistic surveys. Baryon wiggles in the power spectrum correspond
to a single peak in the correlation function. This peak differently
depends on the parameters of the underlying power spectrum,
$\mOmega_{M0}h$, $f_{\rm B}$, and $h$. In the two-dimensional contour
plot of the correlation function in redshift space, the corresponding
feature is the baryonic ridge. The shape of the ridge is perfectly
circular in comoving space, in spite of the effects of the peculiar
velocity. This is preferable for the cosmological test by the AP
effect. Using the analysis of the Fisher information matrix, the
cosmological test by the correlation function in deep redshift space
is shown to constrain the properties of the dark energy component as
well as other cosmological parameters, even if the bias is uncertain.
In particular, there is a clear complementarity of the samples of
various redshift ranges in probing the dark energy component.

Therefore, a survey of a wide redshift range is particularly useful to
probe the dark energy. Keeping a sufficient number density of galaxies
in the given observation time, examples of the optimally designed
survey geometry are illustrated in Figure~\ref{fig5} and
Figure~\ref{fig6}. Selecting galaxies for the spectroscopic follow-ups
as densely as possible is one of the major technical challenges in
this kind of survey strategy. However, recent advances of the
technology enable us to carry it out. With the advent of 8-10m
telescopes, galaxies of $z \simlt 1.4$ can be straightforwardly
selected by color selections as demonstrated by, e.g., the SDSS survey
\citep{york00,eisenstein98} and the DEEP2 Galaxy Redshift Survey
\citep{davis03,coil04}. In $z\simgt 2.5$, Lyman-break galaxies are
photometrically selected with sufficient density
\citep{steidel96,steidel98,adelberger98}. The range $1.4 \simlt z
\simlt 2.5$ is known as the {\em redshift desert}, since spectroscopic
identifications of galaxies were historically difficult. Recently,
\citet{steidel04} showed that large numbers of galaxies in this range
can actually be selected with standard broadband color selection
technique using the Keck I Telescope with the LRIS-B spectrograph.
Thus galaxies at $z\simlt 3$ can be optically selected by ground-based
8-10m telescopes with sufficient number densities
\citep{adelberger04}. Therefore, the proposed surveys illustrated by
Figure~\ref{fig5} and Figure~\ref{fig6} in the redshift range of
$z\simlt 3$ with approximately homogeneous selections are accessible
by present-day technologies.

Cosmology with the large-scale structure has mainly been studied by
the snapshot of the nearly present universe, since the sufficient
number of galaxies have been available only at shallow redshifts, $z
\simlt 0.3$. CMB observations, on the other hand, map the universe of
the $z\simeq 1100$ universe. The observations of the galaxy clustering
in the deep-redshift universe of $0.3 \simlt z \simlt 3$, which are
within reach of the present-day technology, will provide a unique
information on the evolving universe with the analysis of the
correlation function presented in this paper. Observations of the
evolving universe is momentous not only to probe the nature of the
dark energy component of the universe, but also to establish or even
falsify the standard picture of the cosmology. Only data of the
snapshots of the universe are not enough to confirm the consistency of
the cosmological model. Whether or not the structure of the universe
with respect to both space and time are fully explained by minimal set
of cosmological assumptions and parameters will be a major concern in
cosmology. In the era of the precision cosmology, any need for
additional assumptions will suggest terra incognita in the physical
universe.

%\begin{equation}
%\label{eq2-1}
%\end{equation}

%%%%%%%%%%%%%%%%%%%%%%%%%%%%%%%%%%%%%%%%%%

\acknowledgements

I thank Alexander Szalay and Adrian Pope for helpful discussion. I
acknowledge support from the Ministry of Education, Culture, Sports,
Science, and Technology, Grant-in-Aid for Encouragement of Young
Scientists, 15740151, 2003.

%%%%%%%%%%%%%%%%%%%%%%%%%%%%%%%%%%%%%%%%%%

\appendix

\section{Basic Notations}
\label{app0}

In this Appendix, basic notations of the Friedmann-Lema\^{\i}tre
universe with dark energy extension used in this paper are introduced.

The background metric is given by the Robertson-Walker metric,
\begin{equation}
  ds^2 = - dt^2 +
  a^2(t) \left[
    dx^2
    + {S_K}^2(x) \left(d\theta^2 + \sin^2\theta \phi^2\right)
  \right],
\label{eq2-1}
\end{equation}
where we employ a unit system with $c=1$, and adopt a notation,
\begin{equation}
  S_K(x) \equiv
  \left\{
  \begin{array}{ll}
    (-K)^{-1/2} {\rm sinh}\left[(-K)^{1/2} x\right], &
    (K < 0), \\
    x, & (K = 0), \\
    K^{-1/2} \sin\left[K^{1/2} x\right], &
    (K > 0),
  \end{array}
  \right.
\label{eq2-3}
\end{equation}
where $K$ is the spatial curvature of the universe. The comoving
distance $x(z)$ at redshift $z$ is given by
\begin{equation}
   x(z) = \int_0^z  \frac{dz'}{H(z')},
\label{eq2-4}
\end{equation}
where $H(z)$ is the redshift-dependent Hubble parameter. When we allow
the dark energy component having a non-trivial equation of state,
$p(z) = w(z) \rho(z)$, the Hubble parameter is given by
\begin{equation}
   H(z) = H_0
   \left[
      (1+z)^3 \mOmega_{\rm M0} - 
      (1+z)^2 \mOmega_{\rm K0} +
      \exp\left(3\int_0^z \frac{1 + w}{1+z}dz\right) \mOmega_{\rm Q0}
   \right]^{1/2},
\label{eq2-5}
\end{equation}
where $\mOmega_{\rm M0}$ is the density parameter of matter,
$\mOmega_{\rm Q0}$ is the density parameter of dark energy, and
$\mOmega_{\rm K0} = \mOmega_{\rm M0} + \mOmega_{\rm Q0} - 1$ is the
curvature parameter. The linear growth factor $D(z)$ and its
logarithmic derivative, $f(z) \equiv d\ln D/d\ln a = - (1+z)d\ln D/dz$
are the solution of the following simultaneous differential equations
\citep{matsubara03}
\begin{eqnarray}
&&
  \frac{d\ln D}{d\ln a} = f,
\label{eq2-6a}\\
&&
  \frac{df}{d\ln a} = -f^2 - \left(1 - \frac{\mOmega_{\rm M}}{2} -
  \frac{1 + 3 w}{2} \mOmega_{\rm Q} \right) f +
  \frac{3}{2} \mOmega_{\rm M}, 
\label{eq2-6b}
\end{eqnarray}
where the scale factor $a$ is related to the redshift by $a = (1 +
z)^{-1}$, and $\mOmega_{\rm M}$, $\mOmega_{\rm Q}$ are the
time-dependent density parameters of matter and dark energy,
respectively:
\begin{eqnarray}
&&
  \mOmega_{\rm M}(z) = \frac{{H_0}^2}{H^2(z)} (1+z)^3 \mOmega_{\rm M0}
\label{eq2-7a}\\
&&
  \mOmega_{\rm Q}(z) =
  \frac{{H_0}^2}{H^2(z)}
  \exp\left(3\int_0^z \frac{1+w}{1+z} dz\right)
  \mOmega_{\rm Q0}
\label{eq2-7b}
\end{eqnarray}
The normalization $D(z=0) = 1$ is adopted throughout this paper.

\section{Marginalized Fisher Matrix}
\label{app1}

In this Appendix, useful equations to obtain estimates of the error
covariance from the Fisher matrix when some parameters are
marginalized over. The Fisher matrix $F$ is an expectation value of
the curvature matrix of the logarithmic likelihood function in
parameter space,
\begin{equation}
  F_{\alpha\beta}(\bfm{\theta})
  = -
  \left\langle \frac{\partial^2 \ln {\cal L}}
        {\partial\theta_\alpha\partial\theta_\beta}
  \right\rangle,
\label{eqa-1}
\end{equation}
where $\bfm{\theta} = (\theta_1,\theta_2,\ldots)^{\rm T}$ is a set of
model parameters, and ${\cal L}$ is the likelihood function. In the
Fisher matrix analysis, the error covariance matrix $\langle
\Delta\bfm{\theta}\Delta\bfm{\theta}^{\rm T} \rangle$ of a set of
model parameters asymptotically corresponds to the inverse of the
Fisher matrix
\begin{equation}
  F^{-1} \simeq
  \langle\Delta\bfm{\theta}\Delta\bfm{\theta}^{\rm T} \rangle,
\label{eqa-2}
\end{equation}
when all the parameters are simultaneously determined.

The error covariance matrix of partial set of parameters,
marginalizing over other parameters is given by a sub-matrix of the
inverse of the Fisher matrix. Therefore, a Fisher matrix
$\widetilde{F}$ of these partial set of parameters corresponds to the
inverse of the sub-matrix of equation (\ref{eqa-2}). We consider the
situation that the first $n$ parameters are estimated and other $m$
parameters are marginalized over. The full Fisher matrix has a form,
\begin{equation}
  F =
  \left(
    \begin{array}{cc}
      A & B\\
      B^{\rm T} & C\\
    \end{array}
  \right)
\label{eqa-3}
\end{equation}
where $A$ is an $n\times n$ symmetric matrix, $B$ is an $n\times m$
matrix, and $C$ is an $m\times m$ symmetric matrix. A representation
of the inverse of the equation (\ref{eqa-4}) is given by
\begin{equation}
  F^{-1} =
  \left(
    \begin{array}{cc}
      \left(A - B C^{-1} B^{\rm T}\right)^{-1} &
      -\left(A - B C^{-1} B^{\rm T}\right)^{-1} B C^{-1}\\
      -\left(C - B^{\rm T} A^{-1} B\right)^{-1} B^{\rm T} A^{-1} &
      \left(C - B^{\rm T} A^{-1} B\right)^{-1}\\
    \end{array}
  \right),
\label{eqa-4}
\end{equation}
as can be explicitly confirmed that $F^{-1} F = I$. Therefore, the
marginalized $n\times n$ Fisher matrix $\widetilde{F}$ of the reduced
parameters corresponds to
\begin{equation}
  \widetilde{F} = A - B C^{-1} B^{\rm T}.
\label{eqa-5}
\end{equation}
The first term is the Fisher matrix of the reduced parameter space,
without marginalization, and the second term corresponds to the
effects of the marginalization.

Given a Fisher matrix $\widetilde{F}$, with or without
marginalization, the estimations of the error bounds are obtained by
contours of concentration ellipsoid $\Delta\bfm{\theta}^{\rm T}
\widetilde{F} \Delta\bfm{\theta} = {\rm const.}$, where ${\rm const.}$
depends on significance levels and dimensionality of the parameter
space. Therefore, obtaining the marginalized Fisher matrix by the
equation (\ref{eqa-5}) is sufficient to obtain the error estimation by
drawing concentration ellipsoids, and calculation of an inverse of the
full Fisher matrix is not necessary.

Below, we derive the correspondence between contour levels of
$\Delta\bfm{\theta}^{\rm T} \widetilde{F} \Delta\bfm{\theta}$ and the
significance level in the joint estimation of multiple parameters.
Assuming a Gaussian profile around a peak of the likelihood function,
and the correspondence between covariance matrix and the Fisher matrix
of equation (\ref{eqa-2}), an estimate of $\nu \sigma$ error bounds
can be obtained by solving an equation for $X(\nu)$,
\begin{equation}
  \left[(2\pi)^n \det \widetilde{F}^{-1}\right]^{-1/2}
  \int_{\Delta\sbfm{\theta}^{\rm T} \widetilde{F} \Delta\sbfm{\theta} \leq X^2}
  d^n \theta
  \exp(-\frac12\Delta\bfm{\theta}^{\rm T} \widetilde{F} \Delta\bfm{\theta}) =
  \frac{1}{\sqrt{2\pi}} \int_{-\nu}^\nu e^{-t^2/2},
\label{eqa-6}
\end{equation}
where 
\begin{equation}
\Delta\bfm{\theta}^{\rm T} \widetilde{F} \Delta\bfm{\theta} = X^2(\nu)
\label{eqa-7}
\end{equation}
defines a concentration ellipsoid of $\nu \sigma$ significance level.
Adopting a linear transformation of variables, $\bfm{x} = L^{\rm T}
\Delta\bfm{\theta}$, where $L$ is a $n\times n$ matrix by a Cholesky
decomposition of the Fisher matrix, $\widetilde{F} = L L^{\rm T}$, the
equation (\ref{eqa-6}) reduces to a simple equation,
\begin{equation}
  \int_0^X dx x^{n-1} e^{-x^2/2} =
  2^{n/2-1} \mGamma\left( n/2 \right)
  {\rm erf}\left(\nu/\sqrt{2}\right),
\label{eqa-8}
\end{equation}
where ${\rm erf}(x) = 2 \pi^{-1/2}\int_0^x dt e^{-t^2}$ is the error
function. Numerical solutions of this equation in several cases are
given in Table~\ref{taba1}.
\begin{table}[tb]
\begin{center}
\caption{Contour levels $X(\nu)$ of the concentration ellipsoids in
joint estimations of parameters.\label{taba1}}
\begin{tabular}{c|ccc}
\tableline\tableline
 & $n=1$ & $n=2$ & $n=3$\\
\tableline
$\nu=1$ & $1$ & $1.52$ & $2.17$\\
$\nu=2$ & $2$ & $2.49$ & $3.12$\\
$\nu=3$ & $3$ & $3.44$ & $4.03$\\
\tableline
\end{tabular}
\end{center}
\end{table}
In this paper, $1\sigma$ concentration ellipses in marginalized
2-parameter space are presented, i.e., $n=2$, $\nu = 1$ and $X =
1.52$.

%\begin{equation}
%\label{eqa-}
%\end{equation}

%%%%%%%%%%%%%%%%%%%%%%%%%%%%%%%%%%%%%%%%%%

\newpage

%\begin{figure}
%\epsscale{0.95} \plotone{f1.eps} \figcaption[f1.eps]{ caption
%\label{fig1}}
%\end{figure}

\end{document}